\pdfoutput=1
\documentclass[%
 preprint,
 amsmath,amssymb,
 aps,
prl
]{revtex4-2}
\providecommand\JournalTitle[1]{#1}
\usepackage{booktabs}
\usepackage[english]{babel}
\usepackage{graphicx}
\usepackage{dcolumn}
\usepackage{bm}
\usepackage{nicefrac}
\usepackage{hyperref}

\usepackage{soul}
\usepackage[usernames,dvipsnames]{xcolor}
\usepackage{tikz}
\usepackage{comment}

\usepackage[protrusion=false,expansion=true,kerning=true,babel=true,final]{microtype}

\newcommand{\calE}{{\!\!\text{\usefont{U}{calligra}{m}{n}E}\,\,}}

\graphicspath{{Figures/}}

\begin{document}

\preprint{APS/123-QED}

\title{Rapid design of fully soft deployable structures via kirigami cuts and active learning}


\author{Leixin Ma$^{1}$}
\author{Mrunmayi Mungekar$^{1}$}
\author{Vwani Roychowdhury$^{2}$}\email{V.R.: vwani@ee.ucla.edu}
\author{M.K. Jawed$^{1}$}\email{M.K.J.: khalidjm@seas.ucla.edu}

\affiliation{\footnotesize 
$^1$Dept.\ of Mechanical and Aerospace Engineering, University of California, Los Angeles, CA 90095, USA\\
$^2$Dept.\ of Electrical and Computer Engineering, University of California, Los Angeles, CA 90095, USA
}

\begin{abstract}
Soft deployable structures -- unlike conventional piecewise rigid deployables based on hinges and springs -- can assume intricate 3-D shapes, thereby enabling transformative technologies in soft robotics, shape-morphing architecture, and pop-up manufacturing. Their virtually infinite degrees of freedom allow precise control over the final shape. The same enabling high dimensionality, however, poses a challenge for solving the inverse design problem involving this class of structures: to achieve desired 3D structures it typically requires manufacturing technologies with extensive local actuation and control during fabrication, and a trial and error search over a large design space. We address both of these shortcomings by first developing a simplified planar fabrication approach that combines two ingredients: strain mismatch between two layers of a composite shell and kirigami cuts that relieves localized stress. In principle, it is possible to generate targeted 3-D shapes by designing the appropriate kirigami cuts and selecting the right amount of prestretch, thereby eliminating the need for local control. Second, we formulate a data-driven physics-guided framework that reduces the dimensionality of the inverse design problem using autoencoders and efficiently searches through the ``latent" parameter space in an active learning approach. We demonstrate the effectiveness of the rapid design procedure via a range of target shapes, such as peanuts, pringles, flowers, and pyramids. Tabletop experiments are conducted to fabricate the target shapes. Experimental results and numerical predictions from our framework are found to be in good agreement.
\end{abstract}

\pacs{Valid PACS appear here}

\maketitle


Morphing planar geometry to 3D shapes can find a wide variety of engineering applications~\cite{boley2019shape} from additive and subtractive manufacturing to soft actuators~\cite{tang2018design, dias2017kirigami} and architecture~\cite{tomholt2020tunable,panetta2019x,abdelmohsen2019multi}.
Multiple mechanisms have been reported for the 2D to 3D transformation, including residual stress-induced bending~\cite{xu2015assembly}, temperature-induced growth~\cite{kim2012designing}, inflatable membranes~\cite{forte2022inverse}, composite materials controlled by external stimuli (e.g., temperature and pH)~\cite{boley2019shape}, paper folding~\cite{felton2014method}, swelling~\cite{pezzulla2016geometry}, mechanical loads, and boundary conditions~\cite{baek2018form,liu2020tapered,fan2020inverse}.
The fabrication of these shape-morphing structures often requires detailed local control of the geometry, curvature, and stress~\cite{van2017growth, boley2019shape, forte2022inverse, dang2022inverse, wang2021design, xu2015assembly,liu2020tapered,fan2020inverse}. Even though optimizing and realizing arbitrary 3D deformed shapes, such as a human face, is possible, this is often done at a cost of complicated fabrication technique~\cite{boley2019shape}.
Moreover, the inverse design problem of optimizing the physical parameters to achieve the targeted shape typically requires a trial and error search over a high-dimensional space.
To address such shortcomings, this work introduces a new paradigm for planar manufacturing, where a soft kirigami composite (a bilayer shell) deforms from 2D plane to the target 3D shape due to the coupling between two key mechanisms, i.e., kirigami (i.e. material removal)~\cite{moshe2019kirigami, dias2017kirigami,zhang2015mechanically} and strain mismatch~\cite{holmes2019elasticity, stein2019buckling}. Compared to fabrication requiring precise control of various parts of the structure, the reported technique first developed in ~\cite{zavodnik2023soft} -- described later in this paper -- is much easier to implement. 


The convenience of the proposed manufacturing technique, however, is not enough to guarantee adoption unless the associated high-dimensional inverse design problem can also be solved efficiently. In particular, we next address how one can efficiently optimize the structural and geometrical properties to achieve a target 3D shape. 
%
In order to define the problem let us first consider the case without any kirigami cuts. It has been shown that by varying  only the prestretch (but no kirigami cuts), a planar composite plate with strain mismatch  can morph  into different 3D shapes; these shapes, however, are limited to only the few mode shapes~\cite{holmes2019elasticity, zavodnik2023soft}.
To achieve programmability of more classes of shapes one can locally remove materials, thereby introducing kirigami cuts. The optimization problem then reduces to finding kirigami patterns in 2D (i.e. marking the areas where material has to be removed) that would morph the planar structure to a desired 3D shape. However, kirigami cuts have highly nonlocal impact on the structure. A cut at a certain location may affect the global shape of the deformed shell.
Hence, the kirigami pattern optimization for target 3D shapes cannot be conducted locally in space, but requires a global approach to explore the large design space, which is a common problem in many inverse metamaterial designs~\cite{baek2018form,liu2018generative}.

Such inverse design problems where metamaterials are designed for target functionalities  have received significant boost from recent advances in machine learning (ML).  
Various ML algorithms, such as Variational Autoencoder (VAE)~\cite{hanakata2020forward, danhaive2021design} and Generative Adversarial Network~\cite{mao2020designing, liu2018generative} have been successfully applied. However, most of these ML-aided inverse design methods still depend on the generation of a computationally prohibitive number of forward simulation data~\cite{hanakata2018accelerated, mao2020designing}. A network is then trained to learn the inverse map in a supervised manner: the output of the forward simulation is used as input, and the corresponding design parameters as the desired output. This computational overhead of generating forward simulation data is particularly severe for our case. Given a kirigami pattern and other parameters, it takes minutes of computing time to generate corresponding 3D shapes. To train an inverse ML system, one could require millions of such forward simulations, making such inverse approaches computationally intractable for our application. 

In this paper, we make a judicious use of recent advances in dimension reduction techniques (e.g., VAEs), and integrate them with active learning techniques (e.g., Bayesian optimization), where forward computations are done on-demand; that is, only for local explorations.  Such active learning techniques, when used by themselves, typically do not work well for high dimensional search spaces such as in our application. The use of model reduction techniques of ML make these methods practicable.
%
%
%
%
First, classes of candidate kirigami patterns that possess the same symmetry property as the target 3D shape are generated. A VAE is used to parameterize symmetric patterns with low-dimensional latent features and, thus, reduce the dimension of the inverse problem. Then, the optimal latent features are searched iteratively using the Bayesian Optimization framework~\cite{frazier2018tutorial}. 
Numerical simulations and controlled tabletop experiments are conducted to create target 3D shapes with different symmetric properties.
%
%
Excellent agreement is found between experiments and simulations, which demonstrates the prospect of using this method in algorithmic design of metamaterials.

\begin{figure*}[!ht]
\centering
\includegraphics[width=1\columnwidth,keepaspectratio]{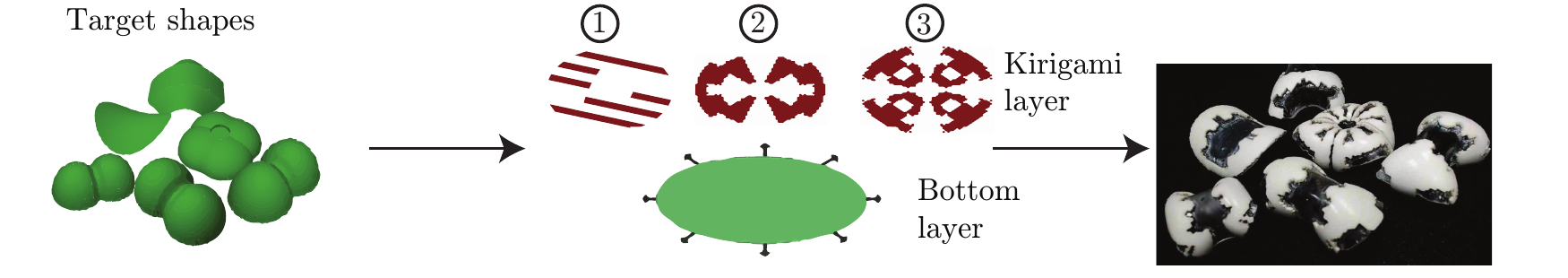}
\caption{The inverse problem aims at designing target 3D shapes by finding the optimal fabrication strategies. Six target shapes are shown on left, while the six experimental results correspond to the target shapes shown on the right. The planar fabrication uses a kirigami layer and a bottom layer is shown in the middle. The three classes of kirigami layers correspond to 1) unidirectional strips 2) reflectional symmetry 3) four-fold radial symmetry. 
}
\label{fig:mlflow1}
\end{figure*}

\begin{figure*}[!ht]
\centering
\includegraphics[width=1\columnwidth,keepaspectratio]{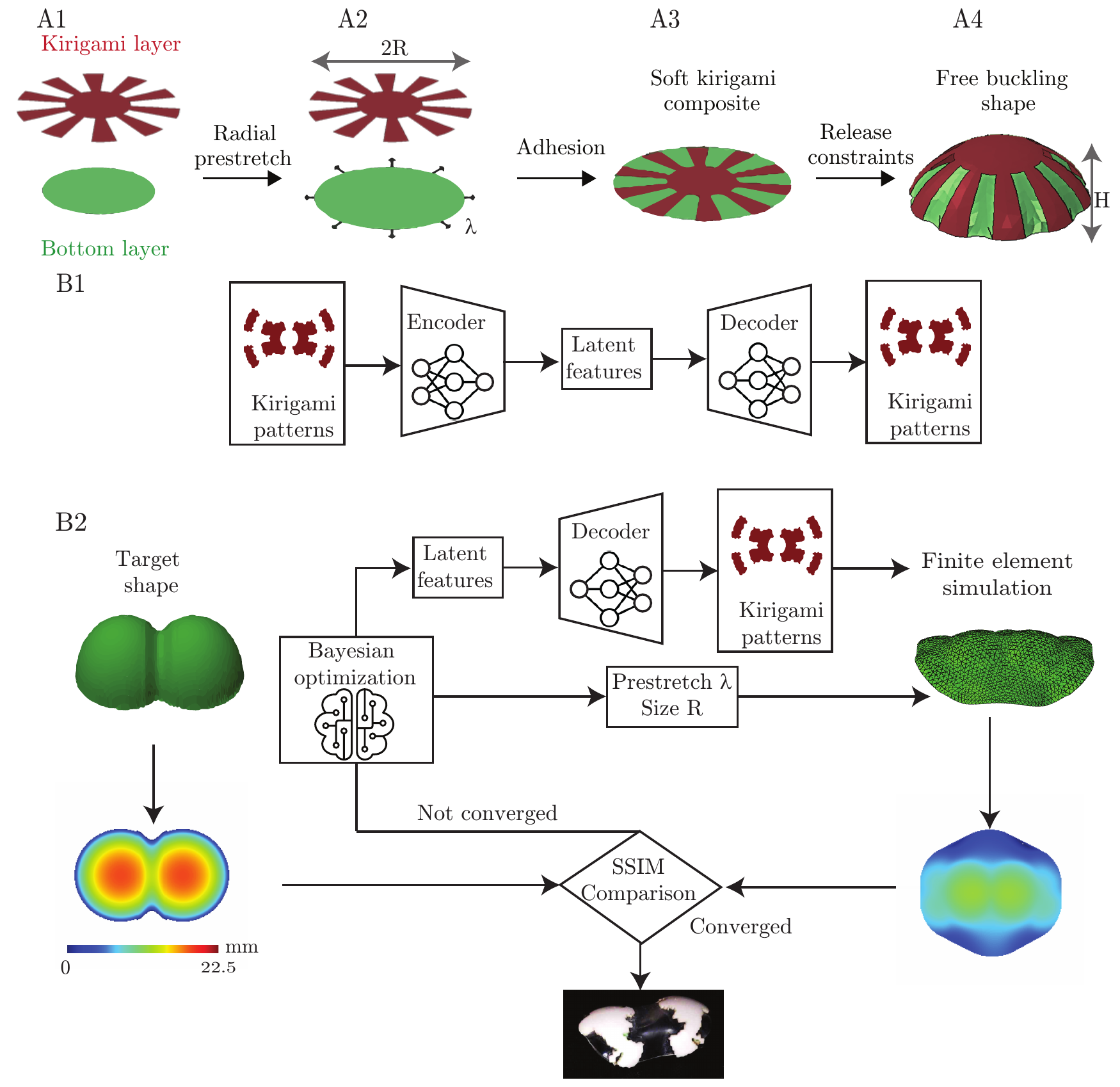}
\caption{A. Overview of the fabrication concept: The composite structure with a kirigami layer and a substrate layer bend to a free buckling shape under radial prestretch. B. Flow chart of the data-driven design and optimization of soft kirigami composite (B1) A VAE to reduce the dimension of kirigami patterns to a small number of latent variables (B2) A Bayesian Optimization loop that iteratively searches the optimal combination of latent kirigami pattern variables, size of the structure, and amount of prestretch that results to the target 3D topology.
}
\label{fig:mlflow}
\end{figure*}

\section*{Overall concept}
\subsection*{Problem description}
The inverse problem aims at designing target 3D shapes by finding the optimal fabrication strategies, as shown in Fig.~\ref{fig:mlflow1}. The proposed fabrication technique is shown in Fig.~\ref{fig:mlflow}{\em A1} to {\em A4}. We start from two thin flexible and stretchable plates (Fig.~\ref{fig:mlflow}{\em A1}). The maximum radius of the top layer is $R$. First, strain mismatch between two plates or layers is created by radially stretching the bottom one with the amount of prestretch $\lambda$ (Fig.~\ref{fig:mlflow}{\em A2}). Then, the top layer of the same radius as the stretched bottom layer is glued onto the bottom layer (Fig.~\ref{fig:mlflow}{\em A3}). The strain mismatch between the two layers induces out-of-plane buckling (Fig.~\ref{fig:mlflow}{\em A4}), since bending is less energetically expensive than compression for thin shells~\cite{pezzulla2016geometry}. However, using strain mismatch in composite shells yields a limited number of structural modes~\cite{holmes2019elasticity, zavodnik2023soft}, and the elastic shells may easily wrinkle (localized deformation) to relax the compressive stresses instead of a global change in shape~\cite{paulsen2016curvature}. Hence, material needs to be strategically removed from the top layer to create kirigami patterns in order to yield a target 3D shape. The theoretical characterization of these kirigami-aided stress relief is still not fully available. The material removal expands the possible number of attainable 3D shapes from 2D, with the out-of-plane buckling depending on both the magnitude of prestretch and geometric parameters (including kirigami cuts)~\cite{shyu2015kirigami,rafsanjani2017buckling}.
Based on this rapid fabrication technique, we aim at inversely designing the optimal kirigami patterns, size of the structure (measured by radius $R$), and prestretch $\lambda$ such that a target 3D shape can be achieved. 

\subsection*{Machine learning framework for design of soft kirigami composite structure}

To tackle the challenges of Edisonian searches, our machine learning-aided design process is illustrated in Fig.~\ref{fig:mlflow}{\em B1} and {\em B2}, and discussed in more detail below.

\subsection*{Dimension reduction: Compact and continuous representation of  kirigami patterns} 



A kirigami pattern can be represented by an $N\times N$ binary image, where the pixels in the uncut areas are represented by 1's. This would correspond to a very high ($N^2$) dimensional search space. Recent advances in Computer Vision (CV), however, have shown that such a family of images typically lies on a much smaller dimensional (say $D$) manifold, referred to as the latent space. One can then search only over a $D$-dimensional space (e.g., $D$=6 in our design), leading to considerable computational savings.  A VAE is one such computational model that can be used to learn a generative model of kiragami patterns, where (a) every kirigami image can be  mapped to a $D$-dimenional vector using an Encoder network, and (b) every $D$-dimensional vector in the latent space can be mapped to a sample kirigami image using a Decoder network. 
The Encoder and Decoder networks are trained simultaneously, such that the latent features are representative enough for various kirigami patterns.  Once trained, any active learning framework can view the space of kirigami patterns as a $D$-dimensional continuous space. This allows one to train the VAE using only a limited set of representative kirigami patterns, and using the VAE to interpolate and generate potentially infinite number of similar kirigami patterns.
The detailed architecture and hyperparameters for the VAE are described in the supplementary material.

\subsection*{Design of numerical experiments}
Optimal design parameters, including the kirigami latent features ($D$-dimensional), size of the structure (measured by radius), and prestretch that yield the target 3D shape were searched by performing iterative Bayesian Optimizations. To search global optimums, the optimization aims to ``explore" (prioritizing regions with large uncertainty) and ``exploit" (focusing on the regions with minimum loss function) the high dimensional parametric spaces, until a suitable loss function is minimized~\cite{burger2020mobile}.
 
To initialize the optimization process, we randomly sampled ten combinations of latent features, sizes of the structure, and prestretches. For each combination of the proposed kirigami patterns and prestretch, numerical experiments were performed via finite element simulations. 
A Gaussian process regression model was constructed to approximate the unknown effect of the design variables on the loss function, which is the error of the 3D shape between the simulations and design target. A common way to represent 3D data is via the projected image of the height ~\cite{ahmed2018survey}. The negative structural similarity index is chosen as the loss function, which is used to characterize the dissimilarity between two images. Compared to mean squared error, the structural similarity index (SSIM) is a perceptually-motivated loss function, which is found to have better performance for image restoration tasks than the squared $l_2$ norm of the error~\cite{zhao2016loss}. 

Once the Gaussian process model is constructed, an expected improvement function~\cite{frazier2018tutorial}, that increases with both the mean of the loss function and its uncertainty, is calculated. Then, the combination of latent features, size of the structure, and prestretch that maximizes the expected improvement function is selected. The corresponding kirigami patterns for that combination of latent features can be recovered from the trained decoder network. The recovered patterns are used to create the mesh for the kirigami layer to perform next round of finite element simulation. The negative SSIM between the results of finite element simulation and target shape is used to update the Gaussian process model and reassess the expected improvement function. Such a process continues iteratively. Once the optimization converges, we verify its effectiveness using precision desktop experiments.

\begin{figure*}[!ht]
\centering
\includegraphics[width=\textwidth,keepaspectratio]{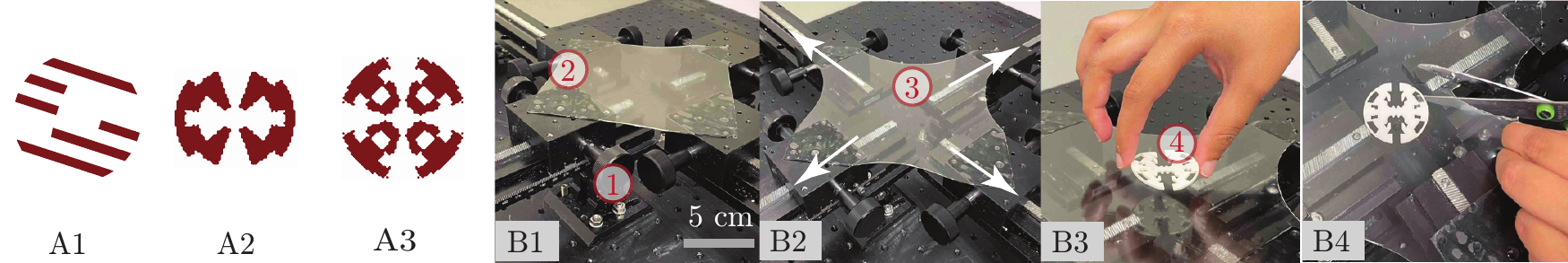}
\caption{A. Examples of different classes of kirigami patterns: 1) unidirectional fibers, 2) shapes of reflectional symmetry with 2 lines of symmetry, 3) four-fold radial symmetry.
B. Experimental setup. (B1) Schematic representation of the system consist of two-knob stages (1)  on 250 mm tracks (2) knobs to adjust the stage location (B2) Snapshot of the system when the substrate layer (3) is stretched (B3) Attach a kirigami layer (4) on top of the substrate layer (B4) Release the prestretch by cutting along the outline of the circular substrate.
}
\label{fig:experiment}
\end{figure*}

\subsection*{Generation of candidate kirigami patterns for VAE training}
We tested the performance of three classes of candidate kirigami patterns, as shown in Fig.~\ref{fig:experiment}{\em A1} to {\em A3}. The first class, as shown in Fig.~\ref{fig:experiment}{\em A1}, is made of uni-directional fibers. For example, Hanakata et al.~\cite{hanakata2020forward} demonstrated that the uni-directional patterns have good expressive power, and can be used to interpolate a variety of mixed kirigami cuts. 
The other classes of kirigami patterns exhibit the same type of symmetric property as the target 3D shapes, as shown in Fig.~\ref{fig:experiment}{\em A2} and {\em A3}. 
We will see later in the paper (Fig.~\ref{fig:allresults}{\em A} and {\em B}) that 
the peanut and pringle like shapes have reflectional symmetry with respect to the two principal axes, while the flower shape has four-fold radial symmetry in Fig.~\ref{fig:allresults}{\em D}.
To create symmetric kirigami patterns, the kirigami cuts are created in certain regions, with the patterns in the remaining regions directly created via rotations or reflections (see \emph{SI Text}).
Both types of non-symmetric and symmetric patterns are augmented via rotation. After rotating the kirigami patterns two to four times, the total number of images for each class is 40955, 40955, and 24573, respectively. These images are used to train VAE models to find low-dimensional representations.

\section*{Physical and numerical Experiments}

\subsection*{Desktop-scale physical experiments}

Fig.~\ref{fig:experiment}{\em B1} to {\em B4} present our experimental setup and the key steps in fabrication. The experimental setup consists of four linear translation stages (250 mm travel, Thorlabs). The substrate and the kirigami layers are made of hyper-elastic materials (see Materials \& Methods). A large substrate layer is applied with the four corners fixed to the four stages (Fig.~\ref{fig:experiment}{\em B1}). Then, to impose the radial strain upon the substrate, we stretch the substrate layer with the same amount in the horizontal and vertical directions (Fig.~\ref{fig:experiment}{\em B}2). Once stretching is complete, a kirigami layer is glued on top of the substrate  (Fig.~\ref{fig:experiment}{\em B3}). We ensure that the central region of the substrate layer -- onto which the kirigami layer is glued -- is uniformly stretched, not affected by the boundaries. 
Afterwards,  Fig.~\ref{fig:experiment}{\em B4} shows that the excess substrate is cut away along the outline of the circular substrate. 
The composite structure then morphs to a certain 3D shape due to the strain mismatch in the two layers. 

\subsection*{Finite element-based numerical simulations}
The finite element software, Abaqus~\cite{abq}, is used to model the nonlinear large deformation of the hyper-elastic composite structures. In the simulations, we use Mooney Rivlin model to simulate the hyper-elastic materials of the substrate and the kirigami layer being tested. 
Each layer of the composite structure was modeled as a shell to reduce the computational cost. The two shell structures are constrained so that their normals match.
The center of the composite structure is fixed.
The prestretch is modeled as isotropic thermal expansion (see {\em SI Text}). The equivalent temperature corresponding to the prestretch is computed. Then, the dynamics of the composite structure (for the given candidate kirigami pattern) is modeled by Abaqus as the 
substrate is cooled back to the reference temperature. The resulting 3D shape is obtained after each such simulation converges. A nonlinear quasi-static simulation is conducted for each design inputs.

The final 3D shape when the active learning search converges will be referred to as the optimal predicted shape; the corresponding kirigami patterns, radius, and prestretch are the predicted optimal parameters.
The detailed structural properties of the composite structure are included in the ``Materials and Methods" section. The numerical simulation is the most time-consuming step, which takes about 1-3 minutes wall clock time on a desktop computer (Ryzen 2950wx CPU @ 2.4 GHz). In total, 100 simulations take about 6 hours. Since forward simulations are expensive, cutting down the number of runs is important.
In this study, we are able to reduce the total number of forward runs from millions to just around 100, while obtaining accurate target shapes. Such computational tractability without compromising performance is one of the key advantages of the proposed machine learning-aided framework.


\section*{Results}
In this section, we test drive the framework with a few example target shapes with different distributions of curvatures and symmetric properties.
%
%
During the optimization, the range of radius of the kirigami structure is constrained between 24 mm and 39 mm while the amount of prestretch $\lambda=1+\epsilon$ is searched between 1.05 to 1.4, which means that the bottom layer is stretched by 5\% to 40\% along both directions. Since in this range of prestretch, the structural property is still properly modeled.


\subsection{Target 3D shapes with reflectional symmetry}
\begin{figure*}[!ht]
\centering
\includegraphics[width=.8\columnwidth,keepaspectratio]{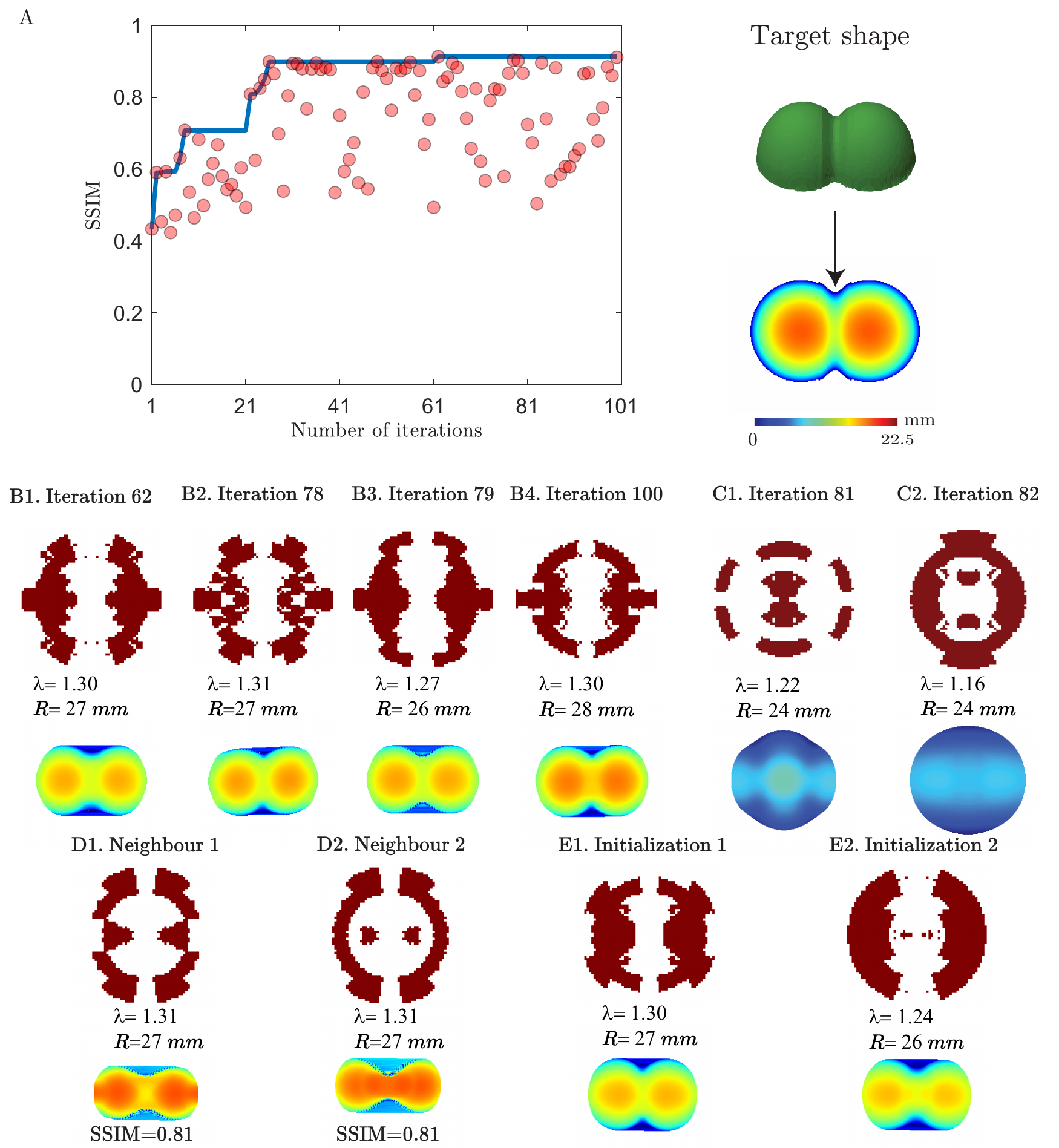}
\caption{\textbf{A. Search trajectory} Variation of SSIM between target peanut shape and predicted free buckling shape over iterations. \textbf{B. Multiple optimal solutions} B1-B4.The top four optimal design solutions, including the kirigami patterns, prestretch and krigami size. The corresponding height distribution for each design is presented at the bottom.  \textbf{C. Exploration during optimization} C1-C2.Two examples showing the algorithm explores new kirigami patterns, that does not lead to closer solutions.   \textbf{D. Interpolated kirigami patterns are often required to better approximate target 3D shapes: } D1-D2. Two nearest neighbors for kirigami patterns found at iteration No. 78, whose free buckling shape gives lower SSIM.  \textbf{E. Effect of initialization} E1-E2. We tested different initialization of the Bayesian Optimization, and presented the results of optimized design parameters and free buckling shape over 100 iterations.}
\label{fig:peanutiter_optimized}
\end{figure*}

As an exemplar of this class, we pick a bilobe structure, resembling a peanut as the target shape. Automatically generated 40,955 (64 by 64) binary kirigami images with reflectional symmetry are used to train the VAE. We find that just a six dimensional latent space is able to reconstruct the training kirigami data with very high accuracy (measured by SSIM $\approx$ 0.97). Thus, the search space of kirigami patterns have been reduced from a $4098$-dimensional binary image space to only a six dimensional continuous search space.
New kirigami patterns can be generated by the VAE, and they are also found to exhibit reflectional symmetry, as demonstrated in the supplementary material. More importantly, as demonstrated in Fig. \ref{fig:peanutiter_optimized} D, the interpolated new kirigami patterns are often needed to generate 3D shapes that best match the target shapes. We next illustrate some of the salient characteristics of our inverse design framework. 

\textbf{Search trajectory in the SSIM space:} The evolution of the design solutions via iterations to create a target peanut shape is demonstrated in Fig.~\ref{fig:peanutiter_optimized}{\em A} and Movie S1.The red dots in Fig.~\ref{fig:peanutiter_optimized}{\em A} plots the variation of SSIM over iterations. 
The maximum SSIMs before certain iteration steps are connected via the blue line. The maximum SSIM improves significantly from 0.7 (in the first 10 random searches) to around 0.91 over time. Some red dots are clustered near the optimal blue line, while some others scatter around the blue line. This suggests that during the Bayesian optimization, the model explores and exploits the design space interchangeably. 
It will not only locally exploit optimal solutions (such as between iteration 78 and 79 in Fig.~\ref{fig:peanutiter_optimized}{\em B}, with materials concentrated to the left and right sides), but also explore unknown design space (such as iteration 81 and 82 in Fig.~\ref{fig:peanutiter_optimized}{\em C}. More materials cover the top and bottom side of the planar structure). 

\textbf{Multiple optimal solutions give SSIM around 0.9:} Fig.~\ref{fig:peanutiter_optimized}{\em B} shows that multiple latent feature combinations can lead to similar 3D deformed shapes. Even though these optimal patterns look different, they share the common characteristics of having more  material concentrated near the left and right ends of the kirigami. The regions with more material compress the bare regions in between where the stiffness is lower, and such a compression leads to a ridge in the middle. 

\textbf{Interpolated kirigami patterns are often required to better approximate target 3D shapes:} We also explored the generalization capability of VAE and the role it plays in optimization. For example, the optimal design shown in Fig.~\ref{fig:peanutiter_optimized}{\em B2} is an interpolated pattern generated by the VAE, and not in the patterns used to train it. We find the nearest neighbours of this optimal generated pattern using K-nearest neighbors algorithm \cite{pedregosa2011scikit}. If we replace the generated pattern with its neighbours in in Fig.~\ref{fig:peanutiter_optimized}{\em D}, we find a decrease in the SSIM. This suggests the importance of interpolation capability in the latent space to achieve more accurate design solutions. 

\textbf{The optimal solutions can be different to the initializations:} We repeat the optimization process several times with different initializations (described in {\em SI Text}). For different initializations that lead to similar values of SSIM (around 0.91) over 100 iterations, the optimized kirigami patterns are different from each other, as presented in Fig.~\ref{fig:peanutiter_optimized}{\em E1} and {\em E2}. However, the optimized prestretch and the radius are close to each other. This suggests that there exists a region of optimal strain-mismatch and structural size that leads to the target 3D shape. 

\begin{figure*}[!ht]
\centering
\includegraphics[width=.8\columnwidth,keepaspectratio]{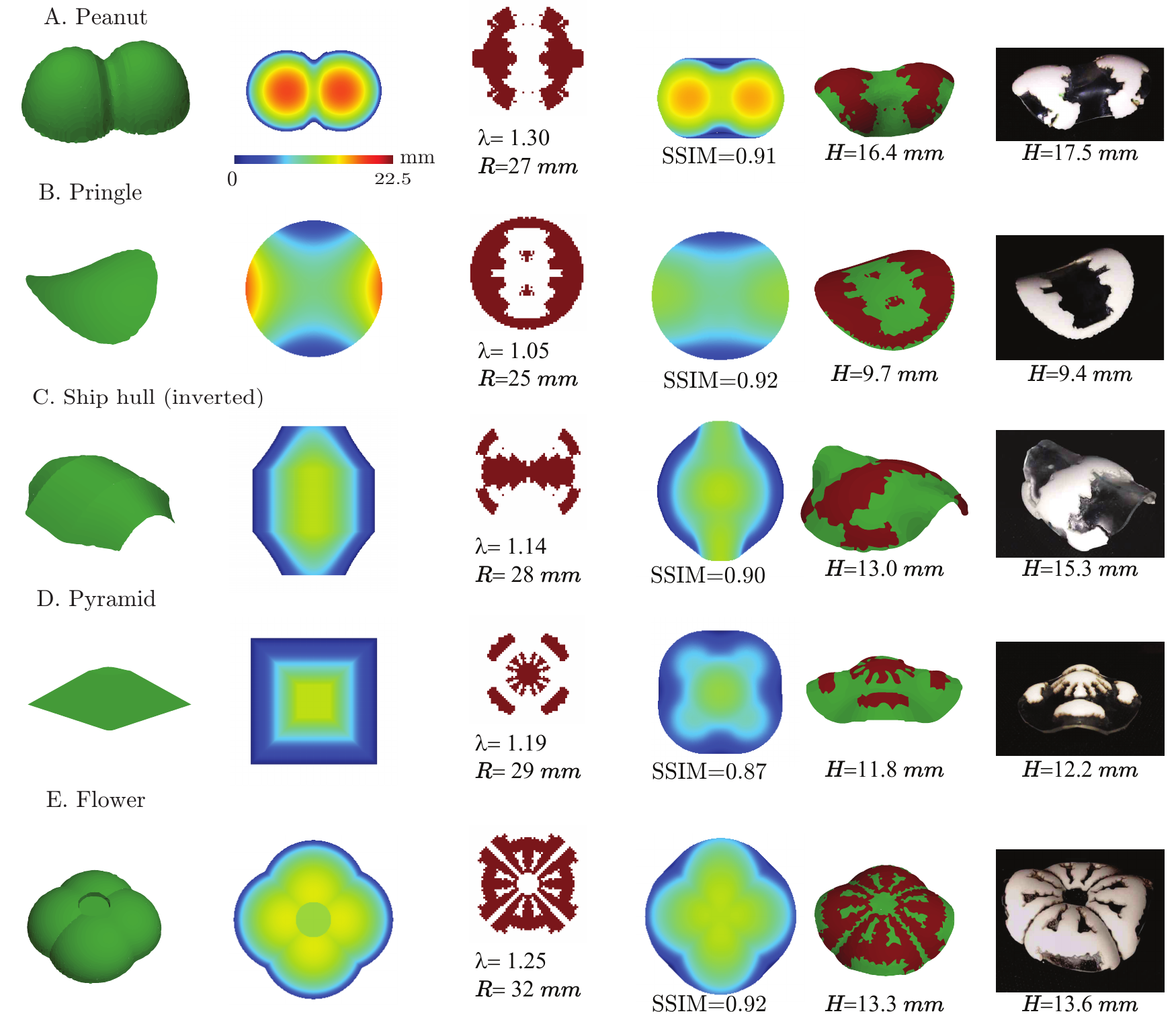}
\caption{\textbf{Inverse design framework: manufacturing programmable soft 3D structures.} We illustrate the end to end design and manufacturing process. \textbf{Target 3D shapes}: the first column shows five different 3D shapes (a peanut, pringle, ship hull, pyramid, flower) input to the optimization algorithm. \textbf{Height-coded 2D image representations}: the second column shows distribution of height for the target 3D shape. \textbf{Optimal design parameters}: the third column presents the optimized kirigami patterns, prestretch, and radius selected after 100 iterations. \textbf{Height-coded 2D image representations of the optimal predicted shapes}: the fourth column presents the height distribution of predicted 3D topology obtained from finite element simulation using the optimal design parameters. The SSIM are approximately in 0.9. \textbf{The predicted 3D shapes}: the fifth column shows the predicted 3D shape in simulation. The maximum heights for the 3D images are shown in the bottom. \textbf{Experimentally realized morphing 3D soft structures with desired shapes}: the sixth column shows the experimental result of the 3D topology given the optimal parameters and following the manufacturing setup illustrated in Fig.~\ref{fig:experiment}}
\label{fig:allresults}
\end{figure*}

We also use the same class of input kirigami patterns to create a pringle and a ship hull-like 3D shape, presented in Fig.~\ref{fig:allresults}{\em B} and {\em C}. The optimized SSIM over 100 iterations is as high as 0.92 and 0.90, respectively. While the maximum SSIM over 10 random searches is at 0.70 and 0.73, respectively. This suggests the importance of strategic search in creating target shapes. The largest errors are found to be near the sharp edges, where the local curvature has sharp changes.

\subsection{Target 3D shapes with four-fold radial symmetry}
Inspired by flowers in nature and man-made pyramids, we further aim to create two corresponding deformed soft structures. Both targets are composed of shapes of four-fold radial symmetry. For example, for each petal in a flower, the structure is bilaterally symmetric, where each half is a mirror image of the other half. The generation of such radially symmetric kirigami patterns are described in {\em SI Text}.
These kirigami patterns can also be represented by six latent features, without sacrificing the high reconstruction accuracy (measured by SSIM $\approx$ 0.99).
%
As shown in Fig.~\ref{fig:allresults}, we get very high SSIM.
For the flower like shape, the maximum SSIM increases from 0.81 in the first 10 random combinations to 0.92 within 100 iterations. The maximum SSIM slightly increases from 0.83 in the first 10 random searches to 0.87 for the pyramid shape.

\subsection{Experimental validation} We carried out the entire end-to-end design process  involving five shapes, shown in Fig.~\ref{fig:allresults}. We picked one optimal design for each target shape from our algorithm. Next we used these optimal design parameters to manufacture the corresponding 3D structures. As Fig.~\ref{fig:allresults} shows, 
the target, the predicted  and the manufactured 3D shapes have the same structural forms, validating our design process. Given such structural similarity,  the maximum height ($H$) of the 3D structure is an easily measured metric for comparison. As shown in the fifth and the sixth columns of Fig.~\ref{fig:allresults}, the $H$ values are in good agreement between the predicted and manufactured 3D structures.  

One can further explore the physics of the kirigami patterns discovered by our design framework. For example, for the four-fold radial symmetry involving the flower and the pyramid target shapes, the optimal kirigami patterns are shown in Fig.~\ref{fig:allresults}{\em E} and {\em D}. The kirigami pattern for the flower shape has materials concentrated in the four lobes and has material removed in the center. This can explain the final shape: the bending of the areas covered with kirigami creates a ridge of slightly lower height in the middle, giving rise to the four lobed flower pattern. In contrast, the pyramid shape with the maximum height in the center requires more material concentrated in the center, as shown in  Fig.~\ref{fig:allresults}{\em D}. 

\begin{figure*}
\centering
\includegraphics[width=1.0\textwidth]{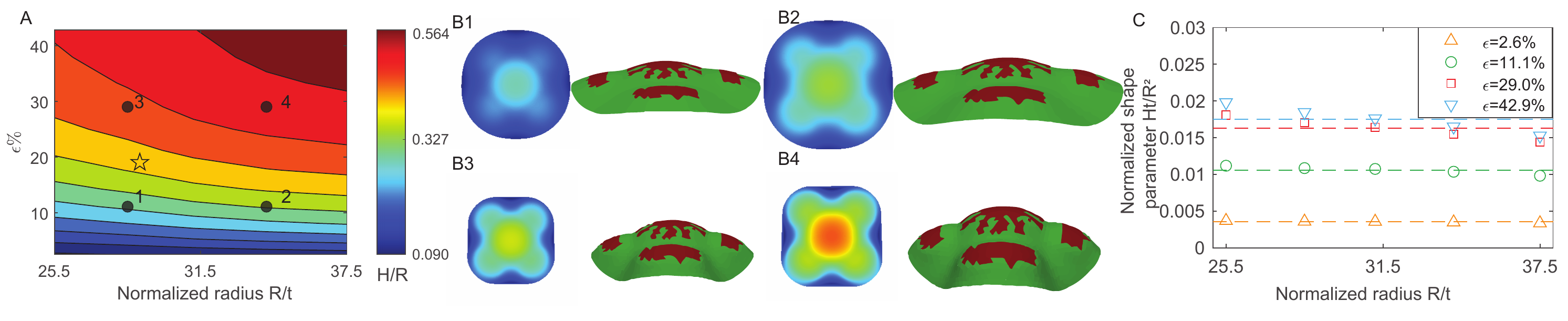}
\caption{\textbf{Scaling law and kirigami invariance}: the same kirigami can be used to design target shapes with varying size parameters $H/R$ by selecting prestretch and radius of the composite structure \textbf{A. Distribution of $H/R$ with the variation of size ratio $R/t$ and the amount of applied prestretch $\epsilon$.} The star symbol denotes the optimal size and prestretch combination that gives the target pyramid shape. The black dots denote the regions where the size and prestretch is perturbed around the optimal point.  \textbf{B. Typical free buckling shapes that correspond to the black dots in A.} Both the height-coded 2D image representation and the corresponding 3D structure are presented. \textbf{C. A normalized shape parameter $Ht/R^2$ as a function of normalized radius $R/t$ at different amount of prestretch.} The line horizontal dashed lines indicates the scaling prediction: $H/R$ scales linearly with $R/t$}
\label{fig:tradeoff}
\end{figure*}

\section*{Discussion}

\subsection*
{Kirigami invariance: Scaling Law} We address the problem of manufacturing the same shape but with different scales (e.g., pyramids with different heights), without having to solve for optimal design parameters for each scale. 
Recall that designing a target 3D shape requires around hundreds of optimization steps, each involving an expensive forward simulation step. Thus, if we can identify a scaling law involving the radius $R$, prestretch $\lambda$, and the scale of the target shapes --while fixing the kirigami patterns-- then such a scaling law can guide designers to quickly vary the scale of the deployable 3D soft structures.

For a given material property and kirigami pattern, the maximum normalized height $H/R$ is affected by prestretch $\lambda=1+\epsilon$, and normalized radius $R/t$, where $H$ is the maximum height of the free buckling shape, and $R$ and $t$ represent the radius and thickness of the planar structure, respectively (see Fig.~\ref{fig:tradeoff}). 
We derive an analytical relationship by balancing the stretching-induced energy (pre-buckling) in the bottom layer and the bending-related energy dominated by the kirigami layer. We find that for a pyramid like 3D shape, the maximum normalized height $H/R$ scales with $\frac{R}{t}\sqrt{\frac{E_{s}}{E_{k}}}{\frac{\epsilon}{1+{\epsilon}}}$ (see {\em SI Text}), where $E_{s}$ and $E_{k}$ are the Young\rq{}s moduli of the substrate and the kirigami, respectively. Such a relationship can directly extend the design of soft composite structures from one maximum height to different heights by changing the original radius of the composite structures and prestretch. 

As an example, we choose the optimized kirigami pattern in Fig.~\ref{fig:allresults}{\em D} that leads to a pyramid shape, and investigate how the normalized height is scaled with the variation of normalized radius and prestretch.
Fig.~\ref{fig:tradeoff}{\em A} shows the maximum normalized height $H/R$ as a function of normalized radius $R/t$ and prestretch. The star symbol in Fig.~\ref{fig:tradeoff}{\em A} denotes the optimal combination of prestretch and radius that leads to the target pyramid shape over 100 iterations. The black dots are the selected examples where the radius and prestretch are perturbed around the optimal values. 
If we want to design taller pyramids, we can slightly increase the normalized radius from Fig.~\ref{fig:tradeoff}{\em B1} to Fig.~\ref{fig:tradeoff}{\em B2}.
When we slightly increase the prestretch around the optimal point from Fig.~\ref{fig:tradeoff}{\em B1} to Fig.~\ref{fig:tradeoff}{\em B3}, the height of the 3D shape also increases, but the radius of the deformed geometry decreases. Similar phenomena can be observed by varying from Fig.~\ref{fig:tradeoff}{\em B2} to Fig.~\ref{fig:tradeoff}{\em B4}. 

The increase of $H/R$ is proportional to the increase of the normalized radius $R/t$ in Fig.~\ref{fig:tradeoff}{\em C}. This agrees with the scaling analysis of elastic energies. 
When the prestretch is larger, the stretching energy in the deformed (i.e. post-buckling) configuration becomes larger, which causes $H/R$ to deviate from the linear relationship with $R/t$. As $R/t$ decreases, the more important the post-buckling stretching energy becomes, but such stretching energy is not considered in the current analytical derivation, which is based on balancing only the pre-buckiling stretching energy.
This suggests that the coupling between stretching and bending are important in these conditions, which needs more complicated models beyond the scaling analysis to explain the coupling and hence the shape variation.
Similar phenomena are also found for other target shapes, such as flower and peanut, and more detailed discussion can be found in {\em SI Text}.


\subsection*{Importance of symmetry}
\begin{figure}
\centering 
\includegraphics[width=.8\linewidth]{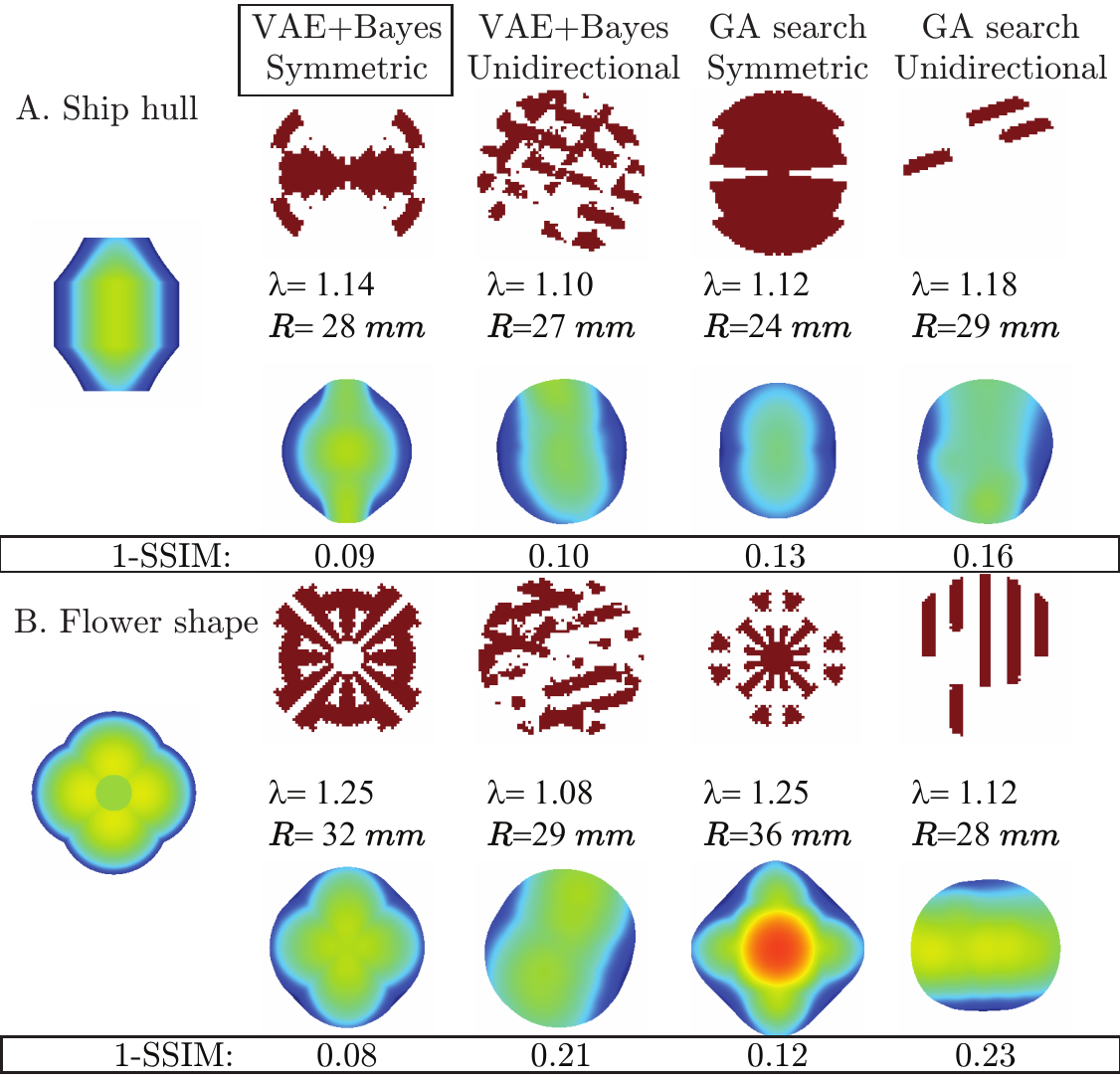}
\caption{\textbf{Comparison of the optimal kirigami patterns and height distribution of assumed 3D shape using different optimization methods.} \textbf{Target shapes:} The first column  shows the distribution of height for the target shapes A. Ship hull B. Flower shape. The optimizations are conducted using four approaches. \textbf{VAE+Bayesian optimization with symmetric patterns:} the second column presents the optimal solution found in 100 iterations using VAE+Bayesian optimization with symmetric patterns. \textbf{VAE+Bayesian optimization with unidirectional strips:} the third column shows the optimal solution in 100 iterations using VAE+Bayesian optimization with unidirectional strips. \textbf{Genetic algorithm with symmetric patterns:} the fourth column shows the optimal solution in 200 iterations using genetic algorithm with symmetric patterns. \textbf{Genetic algorithm with unidirectional strips:} the fifth column shows the optimal solution in 200 iterations using genetic algorithm with unidirectional strips.    }
\label{fig:sup_compare}
\end{figure}

In this section, we aim to demonstrate that an arbitrarily constructed large class of kirigami patterns may still be limited in forming target 3D shapes. The free buckling shape assumed by the symmetry-constrained shapes and unidirectional strips, which doesn't guarantee the reflectional or four-fold radial symmetry, are compared. Fig.~\ref{fig:sup_compare}{\em A} shows that even though the unidirectional strips are able to predict the ship hull like shape with reasonable accuracy, it fails to predict the flower shape in Fig.~\ref{fig:sup_compare}{\em B}. The optimized kirigami patterns and the corresponding 3D shapes after 100 iterations are also compared with the target shape in Fig.~\ref{fig:sup_compare}. Even though the VAE models can generate new patterns, with mixed combination of multi-directional strips, few of the generated patterns are exactly radially symmetric. This causes the free buckling shape to bend more in a preferred direction. The desired reflectional symmetry of a ship hull like target shape is also not guaranteed using the combinations of unidirectional strips. The comparisons demonstrate the importance of constraining the design space with appropriate symmetric property at the very beginning of design process.
\subsection*{The advantage of proposed VAE and Bayesian Optimization combined approach over  evolutionary algorithms}
We also compare the optimization results from the proposed framework with that using genetic algorithm (GA), which is a standard evolutionary algorithm~ (see {\em SI Text}). In the evolutionary approach, we set the variables to be 13 binary integers representing the presence of kirigami materials in the 13 divided regions in Fig.S1 {\em A2}, {\em B2}, and {\em C2} . Another discrete integer is used to indicate the amount of rotation applied to the kirigami patterns, and two other continuous variables are introduced to represent the radius of the structure, and the amount of prestretch to be applied. The population size in the standard algorithm is chosen as 10, and the maximum number of iterations is set as 20. This means that we need to perform 200 numerical simulations per design; recall that this is twice the number of forward simulations used in our design framework, thus giving the GA approach a computational advantage. 

Fig.~\ref{fig:sup_compare} compares the resucolt of applying genetic algorithm to the unidirectional or symmetric shapes.The comparison suggests that the proposed VAE and Bayesian Optimization combined approach can search the optimal combinations of kirigami and prestretch faster, and can achieve  3D deformed shapes closer to target shapes, compared to the evolutionary search. Without the VAE-aided reduced dimensional and \textit{continuous} kirigami search space, the standard evolutionary optimization is limited to searching only the discrete design space. The comparisons between the third and the fifth column of Fig.~\ref{fig:sup_compare}{\em A} and {\em B} demonstrate that the incapability of generating new kirigami patterns beyond the input dataset restricts the conventional optimization approaches from interpolating between the candidate kirigami patterns to achieve desired free buckling shapes.

\section*{Conclusions}

We numerically and experimentally explored the capability of deforming a planar composite structure to target 3D shape via kirigami cutting and strain-mismatch. The design space is very high dimensional to be optimized directly. We formulated a VAE and active learning combined approach to tackle the design challenges. A VAE is used to represent originally high-dimensional design variables to a much lower dimensional continuous search space. The Bayesian optimization is then conducted to quickly obtain multiple optimal design solutions that achieve similar target free buckling shapes, ranging from shapes inspired by a peanut to a pyramid. 
%
We found that the nonlinear interplay between the strain mismatch, size of the composite structure, and the kirigami patterns strongly affect the free buckling shapes.
We also studied the effect of imposing symmetry constraints on the machine learning-aided design results. 
The comparison of the results with and without symmetry constraints demonstrates that the symmetry constraints at the beginning of the machine learning process are important in better approximating the target shapes. A scaling law is used to guide scaling of the target shapes from one size to another, without having to search for the optimal design parameters. We also discussed the advantage of the proposed framework over traditional approaches, such as genetic algorithm. The proposed framework accelerates the design of a series of shape morphing, fully soft composite structures from weeks and months of running millions of simulations to a few hours of strategically examining around 100 examples.

The inverse design method can provide a systematic way to solve a variety of form finding problems not limited to soft kirigami structures, but also to the manufacturing of gridshell~\cite{qin2020genetic,  isvoranu2019x} and compressive buckling-induced 3D architectures using micro ribbons~\cite{luan2021complex}, which can find applications in a range of areas including soft robotics, additive manufacturing, and architecture. In our future work we plan to improve  fabrication accuracy even further by potentially incorporating  controls that can induce local deformations  into our planar-only manufacturing platform. 


\section*{Materials and methods}

\subsection*{Bayesian optimization}

{Bayesian optimization is a sample-efficient approach for solving a wide range of global optimization problems.
See Ref.~\cite{frazier2018tutorial} for details; a short summary follows.
This approach aims to solve an optimization, expressed as 
\begin{equation}
\theta^*=\underset{\theta}{\mathrm{argmax}}f(\theta),
\end{equation}
where $f$ is a black box model which is expensive to evaluate. The function is approximated by a Gaussian process model. A Gaussian process prior using a Matern similarity kernel with homoscedastic noise is selected as the functional prior $\Phi(\theta)$. Given the available data $D$, the posterior distribution of the parameters is computed via the Bayes' theorem,
\begin{equation}
\Phi(\theta)=\frac{\Phi(D|\theta)\Phi_{prior}(\theta)}{\Phi(D)}
\end{equation}
Based on the current posterior distribution, the acquisition function $EI$ is selected as the 
\begin{equation}
EI(\theta)=\mathbb{E}[u(\theta)]=\mathbb{E}[[f(\theta)-f(\theta_t^*)]^+]
\end{equation}
%
%
where $\mathbb{E}$ is computing the expectation, $\theta_t^*$ is the best point observed so far. When $f(\theta)>f(\theta_t^*)$, $u(\theta)=f(\theta)-f(\theta_t^*)$. While when $f(\theta)\leq f(\theta_t^*)$, $f(\theta_t^*)=0$.
The next data to be sampled $\theta_{t+1}$ is selected such that the acquisition function is maximised, i.e., $\theta_{t+1}= \underset{\theta}{\mathrm{argmax}EI(\theta)}$. Then, the Gaussian process and acquisition functions are updated. Such a process iterates several times, until the optimal solution converges.
} The algorithm is implemented using the Scikit-Optimize package~\cite{scikit}.

\subsection*{Implementation details of finite element simulation}
In the finite element simulations, the substrate and kirigami layers were meshed using 3-node triangular shell element. The substrate layer is divided into 2772 triangle elements. The material properties for the substrate and kirigami layer are listed in Table~\ref{table:matparam}.

\begin{table}
\centering
\caption{Material parameters for the substrate and kirigami layer}
\begin{tabular}{lrrr}
Parameter & Value \\
\midrule
$C_1^\textrm{s}$&  2.4 KPa \\
$C_2^\textrm{s}$ & 23.4 KPa\\
Thickness $t^\textrm{s}$ & 1.1 mm\\
$C_1^\textrm{k}$ & -2.6 KPa \\
$C_2^\textrm{k}$ & 185.8 KPa\\
Thickness $t^\textrm{k}$ & 1.4 mm\\
Outer radius of the circular substrate & 3 cm\\
\bottomrule
\label{table:matparam}
\end{tabular}
\end{table}

\section{Data deposition}
We have created a repository at https://github.com/StructuresComp/inverse-kirigami containing the matlab code for generating finite element simulations and python code for performing machine learning.

\section{Acknowledgement}
We thank Shyan Shokrzadeh and Vishal Kackar for their assistance on experiments. Following research grants are gratefully acknowledged: NSF (CMMI-2053971) for L.M., M.M., V.R., and M.K.J.; and NSF (CAREER-2047663, CMMI-2101751) for M.K.J.

\clearpage

\begin{center}
\large{Supplementary Information}
\end{center}

\subsection*{Creating symmetric kirigami patterns}
 For example, when the 3D structure has two symmetry axes,  we divide the 2D structure into four quadrants. According to the symmetric property of target 3D structure, the "planar design space" reduces to the regions in the first quadrants.  In the first quadrant, the kirigami patterns are created by randomly combining the 13 independent pieces, as shown in Figure~\ref{fig:imagerotate} A1, B1 and C1. Then, in case of reflectional symmetry, the kirigami patterns are created by reflecting the image in the first quadrant with respect to the two symmetry axes, as shown in Figure~\ref{fig:imagerotate} B3. 

While in case of structures that are of four-fold radial symmetry, the planar design space further reduces to the half of the first quadrant. The other half can be directly created by the mirror image in Figure~\ref{fig:imagerotate} C3. The final kirigami cuts are created by rotating the patterns of the first quadrants by 90, 180, and 270 degrees, as presented in Figure~\ref{fig:imagerotate} C4. 

\subsection*{Augmenting the kirigami design space via rotation}
\begin{figure}[h!]
\centering
\includegraphics[width=.8\linewidth]{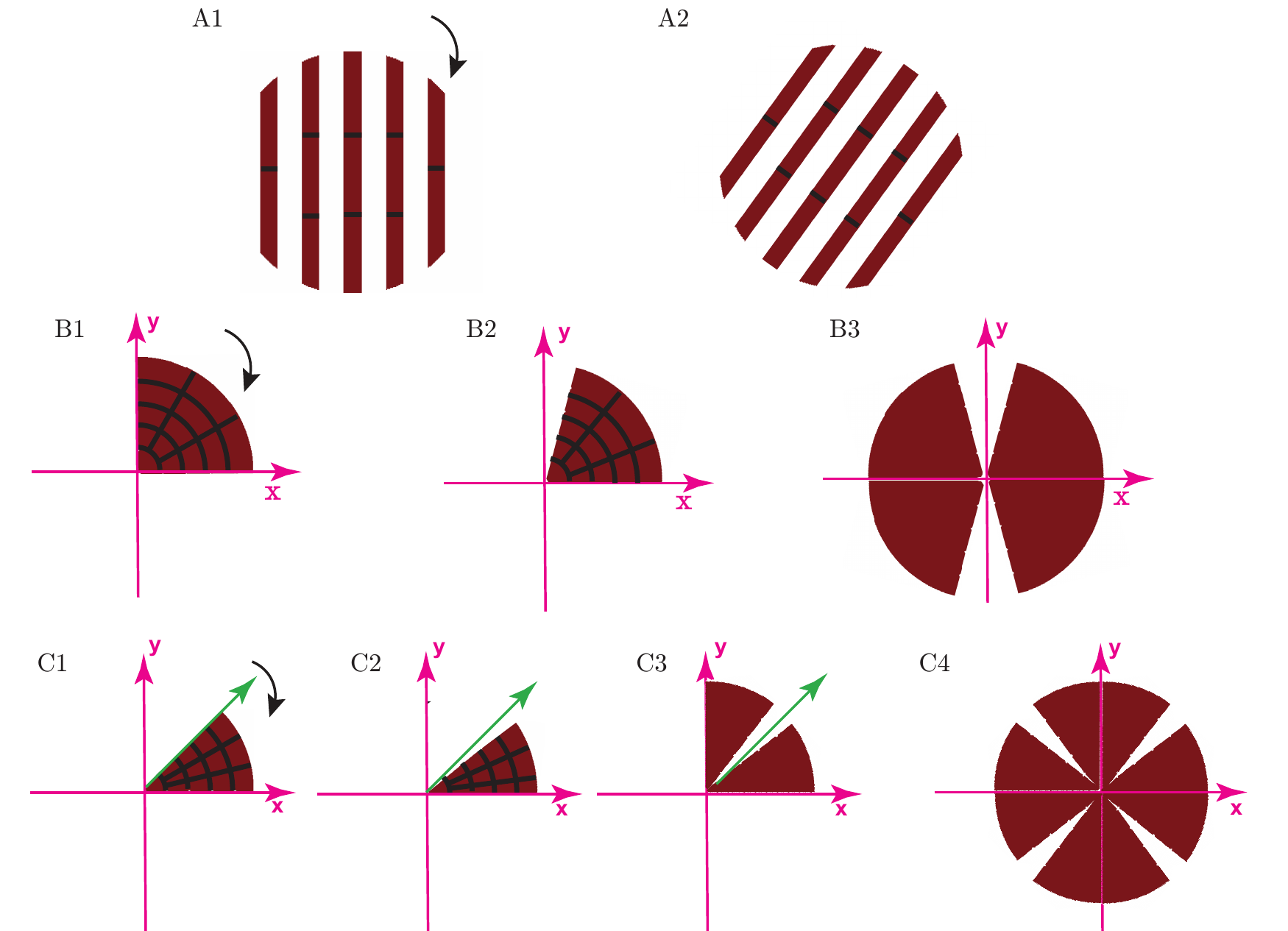}
\caption{Key steps for creating kirigami patterns for A. Unidirectional strips. B. Kirigami patterns that have reflectional symmetry. C. Kirigami patterns that have four-fold radial symmetry }
\label{fig:imagerotate}
\end{figure}

Rotation is a common way to introduce more diverse patterns for machine learning training. 
Once we create a certain set of kirigami cuts in the design space, we further rotate the kirigami cuts in the clockwise and counterclockwise directions two to four times, and only keep the rotated patterns inside the design space, as shown in Figure~\ref{fig:imagerotate} A2, B2, and C2. For the class of kirigami with unidirectional strips, we rotate the strips four times, which are 72, 144, 216, 288 degrees in the clockwise direction. For the class of the kirigami patterns of reflectional symmetry, the kirigami cuts are rotated 7.5 and 15 degrees in the counter-clockwise direction, and are rotated in 4.5 and 12 degrees in the clockwise direction. Finally, for the class of the patterns with four-fold radial symmetry, the cuts in the half of the first quadrants are rotated in 7.5 degrees in the counter-clockwise direction, and in 4.5 degrees in the clockwise direction. 

\subsection*{Variational Autoencoder}
We assume that each observed data point $y_i$ is generated in a nonlinear fashion with latent variable $z_i$, whose joint probability density can be expressed as,
\begin{equation}
p(y,z)=\prod_{i=1}^Np_g(y_i|z_i)p(z_i)
\end{equation}

The full variational lower bound to be optimized include a reconstruction loss (Binary Cross Entropy Loss) and a Kullback-Leibler divergence term, which is written as,

\begin{equation}
ELBO= \mathbb{E}_{z\sim {q(z|y)}}{\log p(y|z)}-D_{KL}(q(z|y)||p(z))
\end{equation}
The approximate posterior given by the inference network for image $y$ is $q(z|y)=N(\mu_z(y),\sigma_z^2(y))$, which follows a normal distribution with mean $\mu_z(y)$ and standard deviation $\sigma_z(y)$. The prior distribution is given as a normal distribution $p(z)=N(0,I)$.

\subsection*{Structure of the Variational AutoEncoder}

The weights in the encoder and decoder are optimized simultaneously to minimize the ELBO loss function, including the KL divergence term and the reconstruction loss. The encoder and decoder each has two hidden layers with 128 neurons per layer. The activation function for each neuron is chosen as the $tanh()$ function. While the activation function for the output is chosen as $sigmoid()$ function. The loss are minimized using the Adam optimization, an extension of Stochastic Gradient Descent with the learning rate set as 0.0005. The input and output for the VAE model are the same binary image, whose size is 64 by 64.


\subsection*{Examples of generated kirigami patterns via the Variational AutoEncoder}

\begin{figure*}[h!]
\centering
\includegraphics[width=.8\linewidth]{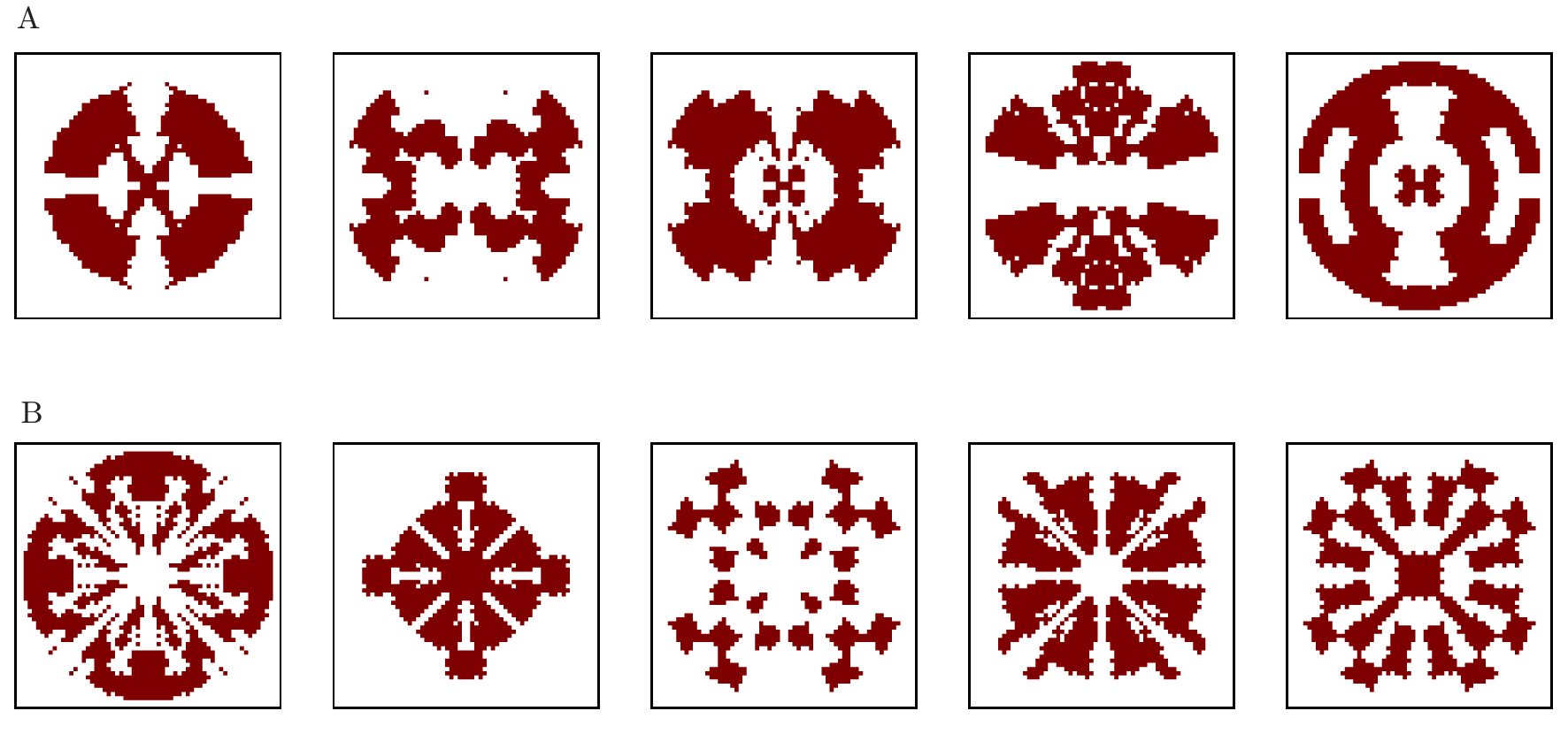}
\caption{Typical kirigami patterns, reconstructed from the 5 cluster centers of the latent features using K-Means clustering A. Reflectional symmetry patterns B. Four-fold radial symmetry patterns}
\label{fig:symclustercenter}
\end{figure*}

Figure~\ref{fig:symclustercenter} presents some typical generated images when images of reflection symmetry (data size of training set is 40955) are input to the Variational Autoencoder model. These images are created by inputting 10 cluster centers in a K-Means clustering process into the trained decoder neural network. It can be found the typical images are also reflection symmetry. 
\subsection*{The effect of initialization on the results for Bayesian optimization}

\begin{figure*}[h!]
\centering
\includegraphics[width=1\linewidth]{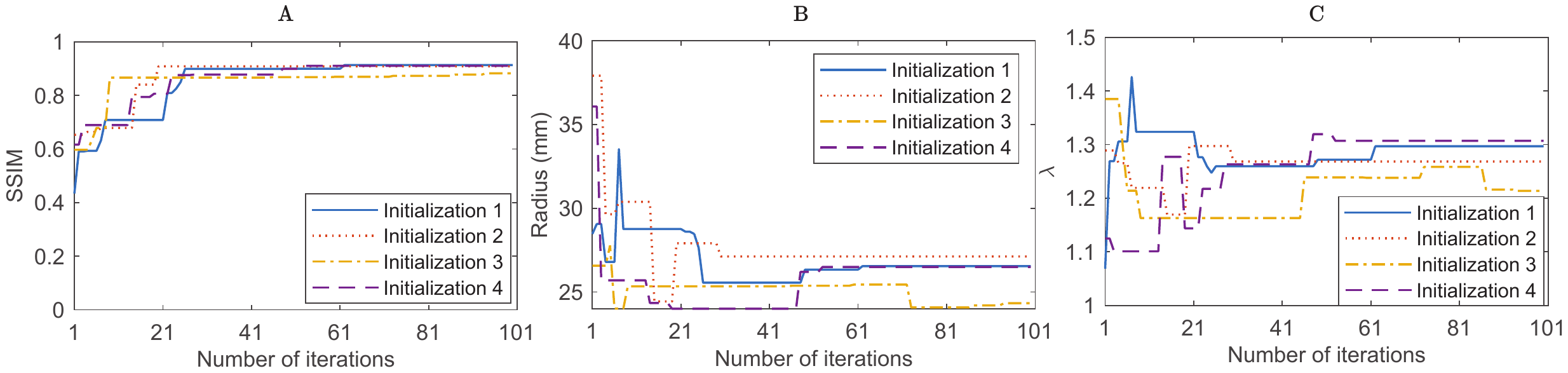}
\caption{For different initalizations, the Bayesian Optimization gives the (A) Variation of maximumm SSIM over iterations.  (B) Variation of kirigami radius over iterations.  (C) Variation of prestretch over iterations (Target shape: peanut shape)}
\label{fig:init}
\end{figure*}

For the peanut shape, we redo the Bayesian optimiaztion with different initial choices of 10 random searches. We find that the initialization affects the maximum SSIM over 100 iterations in Figure~\ref{fig:init}. However, for the initializations that give similar maximum SSIM (Initialization No. 1, 2, 4), the amount of optimized prestretch and kirigami radius are found to be similar, in Figure~\ref{fig:init}{\em B} and {\em C}. This suggests that there exists a sweet-spot of strain-mismatch and structural size that leads to the target 3D shape. 

\subsection*{Detailed description of the genetic algorithm}
\begin{table*}[h!]
\centering
\caption{Parameters for genetic algorithm}
\begin{tabular}{lrrr}
Parameter & Value \\
\midrule
Maximum number of iterations&  10 \\
Population size & 20\\
Mutation probability & 0.1\\
Elite ratio & 0.01 \\
Crossover type & uniform \\
Crossover probability & 0.5\\
Parents portion & 0.3\\
Crossover type & uniform\\
\bottomrule
\label{table:genparam}
\end{tabular}
\end{table*}
The genetic algorithm is a search heuristic relying on biologically inspired operators such as mutation, crossover and selection. 
The idea of selection phase is to select the fittest individuals and let them pass their genes to the next generation.The ``parents'' are selected, based on their fitness scores (e.g. the error between prediction and target). Individuals with high fitness have more chance to be selected as parents for reproduction.
The offsprings are generated by exchanging the genes of the parents at certain probability.These generated offsprings are added into the population. To maintain the diverstiy within the population, a portion of the genes are replaced by a random value. This process is also called as mutation. A pracitcal variant of the general genetic algorithm process also involves allowing the best individuals, determined by the elite ratio,to be directly carry over to the next generation, unaltered. This helps guarantees that the solution quality obtained by the genetic algorithm won't decrease from generation to the next. The algorithm is implemented using \url{https://github.com/rmsolgi/geneticalgorithm}.

When the genetic algorithm is applied to the kirigami composite designs, the inputs (i.e. genes) are variables mixed with integers and continous variables. 
The fitness function to be minimized in the genetic algorithm is the 1-SSIM, where SSIM measures the structural similarity between the prediction and the target. The parameters for the genetic algorithm is listed in Table~\ref{table:genparam}.

\subsection*{Trade-off between size of the composite and the amount of pre-stretch}
\begin{figure*}[h!]
\centering
\includegraphics[width=1\linewidth]{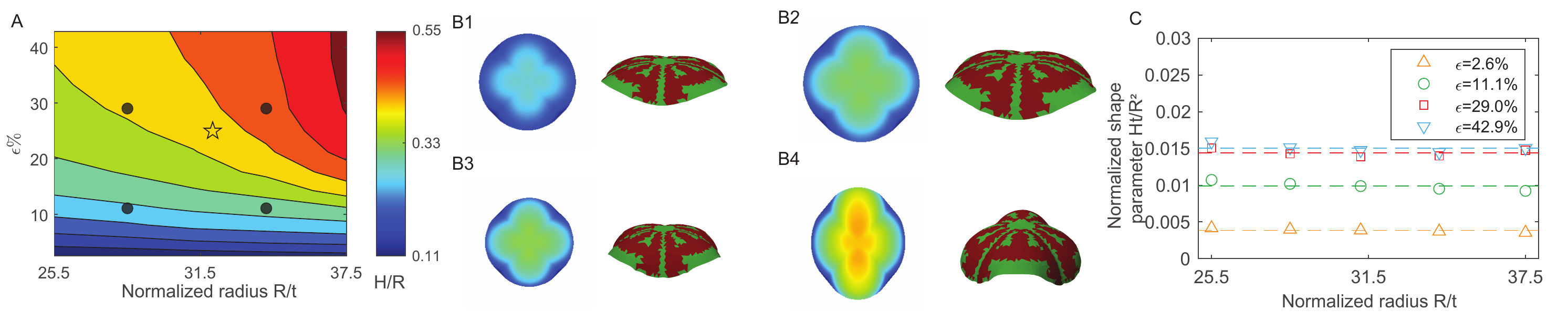}
\caption{(A) Distribution of $H/R$ with the variation of size ratio $R/h$ and the amount of applied prestretch $\epsilon$. The star symbol denotes the optimal size and prestretch combination that gives the target pyramid shape. The black dots denote the regions where the size and prestretch is perturbed around the optimal point.  (B) Typical free buckling shapes that correspond to the black dots in (A). (C) Normalized $Hh/R^2$ as a function of normalized radius $R/h$ at different amount of prestretch. The line horizontal dashed lines indicates the scaling prediction. (Target shape: flower shape)}
\label{fig:tradeoffflower}
\end{figure*}

\begin{figure*}[h!]
\centering
\includegraphics[width=1\linewidth]{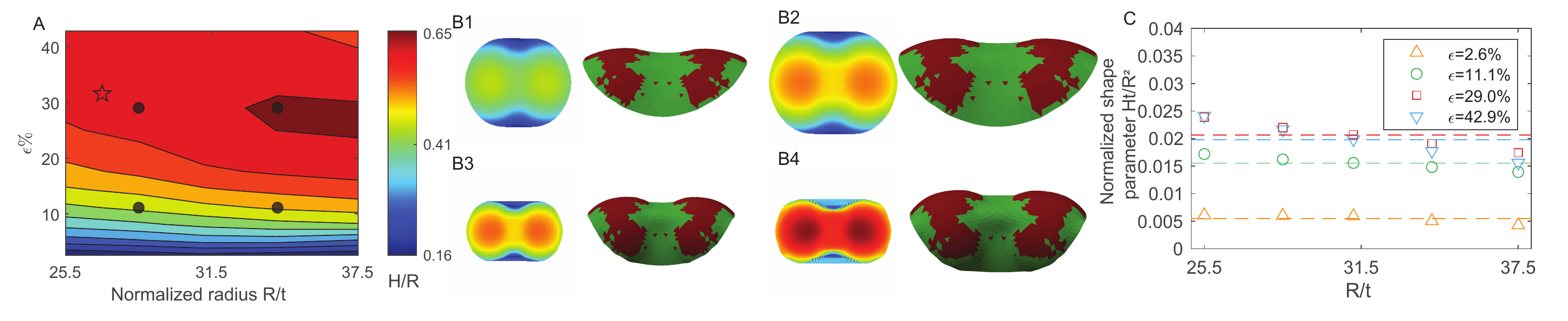}
\caption{(A) Distribution of $H/R$ with the variation of size ratio $R/h$ and the amount of applied prestretch $\epsilon$. The star symbol denotes the optimal size and prestretch combination that gives the target pyramid shape. The black dots denote the regions where the size and prestretch is perturbed around the optimal point.  (B) Typical free buckling shapes that correspond to the black dots in (A). (C) Normalized $Hh/R^2$ as a function of normalized radius $R/h$ at different amount of prestretch. The line horizontal dashed lines indicates the scaling prediction. (Target shape: peanut shape)}
\label{fig:tradeoffpeanut}
\end{figure*}

If we keep the kirigami fixed, the target shapes of other 3D geometries also have nonlinear relationships with prestretch and kirigami size, especially when prestretches are large, as shown in \ref{fig:tradeoffflower} and \ref{fig:tradeoffpeanut}. In general, we can see that the maximum height $H/R$ increases together with the prestretch and structure size. The radius of the deformed shape increases with the structure size, but decreases with larger prestretch. Even though the linear relationship between $H/R$ and $R/h$ holds well for small prestretch conditions, as prestretch increases, the 3D shape can be very different from the deformation under small prestretch. The actual relation between $H/R$ and $R/h$ starts deviating from the linear relationship. For instance, for the kirigami pattern that creates the targeted peanut shape, if we decrease the prestretch to a small enough value, the 3D shape also turns into a pringle shape, rather than maintaining the peanut shape (\ref{fig:tradeoffpeanut}). 

\subsection*{Simulation of prestretch release via temperature reduction }
To simulate the prestretch release, we apply an isotropic expansion. 
\begin{equation}
    \begin{aligned}
        \epsilon_{sim}=\alpha(\theta-\theta^0)=\alpha\delta \theta
    \end{aligned}
\end{equation}
where $\alpha$ is the thermal expansion coefficient, which is set to 1, $\epsilon_{sim}$ is the strain induced by the temperature variation. $\theta$ and $\theta_0$ are the current and initial temperature, respectively.
The temperature variation $\delta \theta$ is relatd to the prestretch in the experiments $\epsilon$ via
\begin{equation}
    \begin{aligned}
        \epsilon=\frac {\delta \theta}{1-\delta \theta}
    \end{aligned}
\end{equation}


\subsection*{Analytical analysis of the effect of design parameters on the free buckling shape}

For small strain, we calculate the stretching-induced energy for the substrate layer as,
\begin{equation}
    \begin{aligned}
        \calE_s=\frac{1}{2}{\pi {R^2}}C_s\delta \theta^2
    \end{aligned}
\end{equation}
where $C_s= E_st/(1-\nu^2)$.$t$ is the thickness of the plate, which is around $1 mm$ in the experiment.

We assume that after the deformation, the structure is bending dominated, and the curvature is about the same everywhere. The bending energy can be calculated as,
\begin{equation}
    \begin{aligned}
        \calE_b &= \frac{1}{2}{\pi {R^2}}D_{s}\kappa^2 + \frac{1}{2}n{\pi {R^2}}D_{k}\kappa^2
\approx \frac{1}{2}n\pi {R^2}D_{k}\kappa^2
    \end{aligned}
\end{equation}
where $n$ is the fraction of the area covered by the kirigami layer, which is a factor between 0 and 1. $D_{s} = \frac{{E_{s}{t^3}}}{{12\left( {1 - {\nu ^2}} \right)}}$ and $D_{k} = \frac{{E_{k}{t^3}}}{{12\left( {1 - {\nu ^2}} \right)}}$ are the bending rigidity of the substrate and kirigami layer, respectively. 

\begin{figure*}[h!]
\centering
\includegraphics[width=1\linewidth]{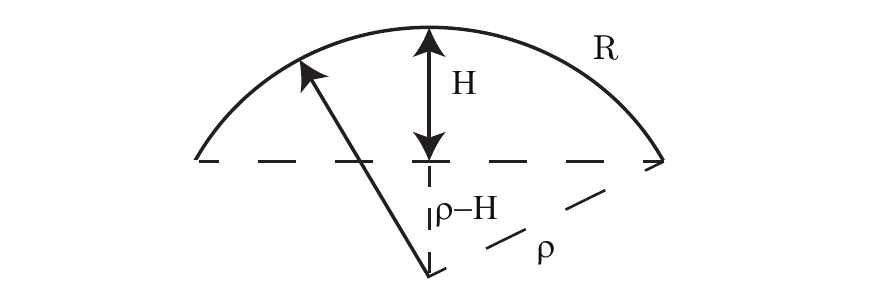}
\caption{Sketch of a spherical cap with height $H$}
 
\label{fig:ana}
\end{figure*}

We further assume that the deformed shape is close to a spherical cap shape with maximum height $H$, and has approximately constant curvature everywhere, as shown in Figure~\ref{fig:ana}.
The length of the curve $R$ can be approximated as, 

\begin{equation}
    \begin{aligned}
            R^2= \rho^2-(\rho-H)^2+H^2
    \end{aligned}
\end{equation}
where $\rho$ is the radius of the hemisphere, which is related to the curvature $\kappa$ via $\kappa= 1/\rho$.

Hence, the maximum height of the shape is related to the curvature via,
\begin{equation}
    \begin{aligned}
        H= \frac{R^2}{2}\kappa
    \end{aligned}
\end{equation}

Hence, equalizing the bending energy and stretching energy, and combining equation (5) gives,
\begin{equation}
    \begin{aligned}
        \frac{H}{R}=\sqrt{\frac{3E_s}{nE_k}}\delta \theta\frac{R}{t}
        \approx \sqrt{\frac{3E_s}{nE_k}}{\frac{\epsilon}{1+{\epsilon}}}\frac{R}{t}
    \end{aligned}
\end{equation}

\begin{equation}
    \begin{aligned}
        \frac{H}{R}
        \approx \sqrt{\frac{3E_s}{nE_k}}{\frac{\epsilon}{1+{\epsilon}}}\frac{R}{t}
    \end{aligned}
\end{equation}

\end{document}